\documentclass[10pt,letterpaper]{article}

\usepackage{iccm}

\iccmfinalcopy 

\usepackage[T1]{fontenc}
\usepackage{pslatex}
\usepackage{listings}
\usepackage[colorlinks=true, allcolors=blue, citecolor=black]{hyperref}
\usepackage{apacite}
\usepackage{color}
\usepackage{graphicx}
\usepackage{amsmath}
\usepackage{float} 

\definecolor{codegray}{rgb}{0.5,0.5,0.5}
\definecolor{codebackground}{rgb}{0.95,0.95,0.95}
\definecolor{promptcolor}{rgb}{0.1,0.1,0.6}

\title{Rapid Prototyping of Event-Driven Contextual Memory \\ in the ACT-Up Cognitive Architecture}
 
\author{{\large \bf Robert Thomson (thomsonr@cmu.edu)} \\
  Department of Psychology, Carnegie Mellon University \\
  Pittsburgh, PA 12513 USA\\
  \AND {\large \bf Christian Lebiere (cl@cmu.edu)} \\
  Department of Psychology, Carnegie Mellon University \\
  Pittsburgh, PA 12513 USA}

\begin{document}

\maketitle

\begin{abstract}
The present paper describes an implementation of contextual memory and a basic event-handler for the ACT-Up cognitive architecture which maintains its scalability and appropriateness for rapid-prototyping while adding essential features and lowering the barrier to entry for new users. This includes describing a theory-neutral implementation of working memory and spreading activation, in addition to a basic associative learning mechanism. An example of rapid prototyping for algorithm development is presented using the serial memory task described in \citeA{klein2005comparative}. This study describes how contiguity effects change across sequential list presentations across three serial and free recall conditions. We further describe how to use generative AI and the event handler to automatically create cognitive experiments directly from the Methods section of research papers. 

\end{abstract}

\section{Background}
In the past decade there have been several advances modeling statistical (i.e., \textit{associative}) learning in cognitive architectures. \citeA{thomson2015account} and \citeA{thomson2017account} presented an interference-based associative learning mechanism, while \citeA{anderson2020associative} presented an updated Bayesian mechanism that does not fall prey to the instabilities of prior implementations (see \citeNP{thomson2013constraining} for a review), and \citeA{kelly2015holographic} described a holographic memory implementation of ACT-R's declarative memory which captures many features of associative learning.

One thing common to these previous implementations is the commitment to some version of the full ACT-R cognitive architecture, including major commitments such as persistent \textit{buffers} to contain information and a \textit{production system} to trigger processing. One issue with these previous implementations is that scalability becomes an issue when trying to integrate these mechanisms into larger and longer-running models \cite{stocco2024integrated}. To address issues of scalability and rapid prototyping, ACT-Up \cite{reitter2010accountable} was developed to model the core functionality of declarative memory with minimal commitments to other theoretical and implementation-level constraints (e.g., there are no buffers, no automatic procedural system, and no built-in sensory mechanisms to receive input). Instead, the implementation details are left to the modeler while maximizing the scalability of the declarative memory system \cite{lebiere2015functional}. More recently, a Python implementation, PyACT-Up \cite{yang2020pyactup} was developed to further integrate with modern AI development.

While ACT-Up (and PyACT-Up) offer flexibility, there are several core features missing which limit their ability to capture human performance on par with the complete ACT-R architecture. These key limitations include: 1) a lack of spreading activation, 2) a limited mechanism to capture information flow within the model (e.g., contextual memory), and 3) no simple event-handling to model the flow of human experiments. As such, while the strengths of scalability and rapid prototyping are substantial, the lack of user-friendly features is challenging for new researchers. To address these shortcomings, the present paper describes a basic, scalable implementation of contextual memory (working memory with associative learning) and an event-handling system for ACT-Up which provides a more complete (yet still scalable) rapid-prototyping framework for developing models and algorithms.

\section{Contextual Memory Module}
Contextual memory can be best described as the conjunction of working memory and associative learning that integrates mechanisms for priming and interference in working memory \cite{thomson2014extending}. It has been previously argued that a limitation of contextual memory in the standard ACT-R architecture is that the commitment to buffers is an \textit{all-or-nothing} situation where a chunk is either in working memory and a possible source of activation, or it is not. There is no default temporal \textit{decay} or interference-based attenuation of potential sources of activation (the $w$ parameter in canonical ACT-R). While extant modules have been described previously \cite{thomson2014extending, thomson2015account, thomson2017account}, they are still generally bound by the commitment that a chunk held within an ACT-R buffer should maximally be a source of activation, even if there is the addition of a decaying context window. 

Due to issues of scalability and lack of stable associative learning (at the time) in the full ACT-R architecture, neither spreading activation nor associative learning mechanisms were developed for ACT-Up. As more projects require human-AI collaboration, the requirement to maintain contextual/conceptual alignment between human operators and AI models has necessitated more functionality than the core memory functionality (partial matching and blended retrievals) can provide. Furthermore, a study by \citeA{lupyan2026unreasonableeffectivenesspatternmatching} demonstrates the importance of similarity, context, and pattern matching to explain human and AI behavior.

\subsection{Working Memory API}
This presents the main functions supporting working memory functionality within the context memory package. The goal is to provide a theory-neutral implementation of working memory where elements are added and removed from context to be available for spreading activation, associative learning, or other mechanisms (e.g., analogical transfer). 

\subsubsection{init-context-module () | reset-context ()} Initialize | reset the context-module. Module must be initialized before use.

\subsubsection{add-element (element) | remove-element (element) | update-element (element)} Adds | removes | updates a context-element from working memory. 

\subsubsection{update-context (current-time  update-function threshold)} Updates all elements in working memory based on a user-supplied \textit{update-function}. The default uses ACT-R's activation as described in \citeA{thomson2014extending} and seen in Equation \ref{eq:base-level}, with an optional \textit{threshold} parameter stating at what residual activation in working memory to remove elements.

\subsection{Associative Learning API} 
This presents the main functions supporting associative learning within contextual memory. This functionality stores associations in a hash-table allowing for constant-time access to update sources of activation, providing scalable learning. Functionality that controls associative growth, spreading activation, and interference effects is provided by user-defined functions. We also include several prototype functions as described below. While we assume that associations do not decay temporally, this is possible to implement. As described in \citeA{thomson2013constraining}, it is generally not recommended to have both associations and chunk activations temporally decay. This is essential to maintain stable spreading activation over longer durations. It is possible to replace the \textit{update-association} function to substitute an entirely different associative learning framework.

\subsubsection{add-association (source target count)} Adds a directed association from \textit{source} to \textit{target} with initial value \textit{count}. 

\subsubsection{update-association (source target new-count \&optional interference-function association-function context)} Updates an existing association to a new value \textit{new-count}.

\subsubsection{update-from-context (target)} Creates and/or updates associations between items in working memory and a given \textit{target}, usually after a retrieval, although this is up to the user. 

\subsubsection{spreading (target context)} Returns the total spread from all sources in $context$ (working memory) based on their residual activation and associative strength. 

\subsection{Prototype Associative Spreading Functions}
The following functions describe a basic associative learning (capped-spread) and two candidate working memory decay functions: one based on canonical ACT-R and the other adding interference based on the number of items in working memory. The adaptive interference provides a soft-cap for the number of elements possible in working memory. Using the default decay of 0, it is possible to programmatically set the contents of working memory (i.e., the \textit{context}) entirely driven by the user. The current example provides an associative learning mechanism that stores raw contextual activation and uses \textit{capped-spread} to provide a shifted hyperbolic transformation function to normalize activations within $[0,1]$ with growth and interference. 

\subsubsection{capped-spread (freq \&optional (max 1) (b 0.5))} Scales the output of $spreading * [0, max]$ using the following function: $max*(tanh(b*freq))$ which provides constrained growth. This function may be substituted to prototype other candidate spreading functions and is referenced using the $*spreading-function*$ parameter.

\subsubsection{actr-decay-activation (current-time add-time decay .35)} Uses a simplified form of base-level decay to reduce the activation of chunks in working memory: \begin{equation} 1 - d*\log\left(1 + (t - t_{0})\right) \end{equation} where \textit{d} is the decay, \textit{t} is the current time, and $t_{0}$ is the time the element was added to working memory. Current default \textit{decay = .35} as argued in \citeA{thomson2014extending}, being approximately 18s for an unrehearsed element to maintain activation, similar to that shown in classic memory experiments \cite{peterson1959short}.

\subsubsection{adaptive-decay-activation (current-time add-time w)} Adapts the simplified form of base-level decay to reduce the activation of chunks in working memory integrating interference by increasing the decay rate based on the number of elements in working memory: \begin{equation}1 - \frac{\log\left(|context|
\right)}{w} *\log\bigl(1 + (t - t_{0})\bigr)\end{equation} where \textit{w} determines how much the decay rate increases based on the number of elements in working memory. 

\subsubsection{use-based-decay (value interference time)} Interference function that takes an existing associative \textit{value} and attenuates it based on a supplied \textit{interference} amount. The current function simply attenuates by a parameterized percentage (set to 1\%) of the total current associative strength. The \textit{time} is an optional parameter for those who want to evaluate temporal-based decay for associations. 

\subsection{Helper Function API}
This presents some helper functions to help modelers view and match associations.

\subsubsection{get-matching-elements (list-slots-test)} Used to search working memory using user-supplied function; e.g., \textit{(activation .5 >=)} or \textit{(element 'cat' equalp)}. 

\subsubsection{pprint-source-associations | pprint-source-associations-tabular} Prints out list of associations for debugging.  
\vspace{6pt}

\section{Event Handling} 
While the ACT-R architecture has a complete event-handling system to handle stimuli entering the sensory buffers (e.g., adding items to the visicon or aural buffers), ACT-Up was designed to be integrated into real-world examples; however, this requires users to develop their own systems for simulating human studies. We developed a scalable, bare-bones event system for simulating external input like that seen in cognitive psychology experiments (\href{https://github.com/cmu-psych-fms/event}{https://github.com/cmu-psych-fms/event}). While we use the term \textit{event buffer} to refer to the contents of the event system, we are agnostic to the structure of the items (i.e., we make no commitments whether they are strings, symbols, or chunks) and have no commitment outside of it being a structure to hold information for the model to perceive. A brief API of the system is as follows:

\subsubsection{reset-event-buffer (reset-clock)} Initializes the \textit{event buffer}, \textit{event queue}, and \textit{event log}. Optionally resets the event clock.

\subsubsection{queue-item (item onset-time offset-time)} Queues adding an \textit{item} to the \textit{event buffer} with a given onset and offset.

\subsubsection{show-item (item duration)} Adds an \textit{item} directly to the \textit{event buffer} for a given \textit{duration} in seconds.

\subsubsection{purge-item (item event-time)} Removes an \textit{item} in the \textit{event buffer} at a given \textit{event-time}.

\subsubsection{advance-event-clock (time)} Advances clock by \textit{time}, processing all events and logging exact onsets and offsets.

\subsubsection{current-event-items} Displays items in the \textit{event buffer}.

\subsubsection{event-log} Returns a history as a list of dictionary elements with keys \textcolor{red} {:time} , \textcolor{red} {:event}, and \textcolor{red} {:item}. \\

This API is sufficient to develop single- and multi-stimuli cognitive psychology studies and can be synchronized with act-up or act-r time. Furthermore, using Perplexity (or similar) generative AI applications, we were able to attach our event.lisp file and the \citeA{klein2005comparative} PDF to accurately create a skeleton Lisp function that completed all 3 conditions of the study using the prompt in Listing \ref{lst:sample_code}.

\label{listing}
\begin{lstlisting}[language=bash, label={lst:sample_code}, caption=AI Prompt to create the serial memory study.]
Implement the attached study using the event handler described in the event.lisp file. 
\end{lstlisting}

\section{Spacing Effect and Base-Level Activation}
The base-level activation of memories reflects the timing of their presentations and rehearsals according to an equation capturing the power laws of practice and forgetting \cite{anderson1991reflections}. While that learning mechanism captures many long-term trends accurately, it fails to reflect finer-grained temporal patterns such as the spacing effect, whereby massed presentations are less well recalled than when more evenly spaced out. \citeA{pavlik2005practice} proposed an account of the spacing effect that varied the decay parameter of each memory presentation according to its activation at the time. An alternative account has been proposed where each presentation was associated with a weight that aimed to boost the base-level activation back to a pre-specified level. A practical advantage of this approach is to prevent unbounded activations resulting from out-of-control reinforcement loops. Specifically, the base-level activation of a memory is defined as:

\begin{equation}
\label{eq:base-level}
B_i=\log(\sum_{j=1}^{n}w_j * t_j^{-d})
\end{equation} 

\noindent where $w_j$ is defined such as $B_i=c$ at $t$ time units in the future, with $c$ as the target base-level activation and $t$ is a short-term horizon to allow for some decay of the most recent activation boost. This adaptation of the spacing-effect has been integrated into the present model.

\section{Case Study: Serial Memory}
The present study reproduces the serial memory experiment described in \citeA{klein2005comparative}. Whereas most serial memory studies employ a single list presentation, \citeA{klein2005comparative} presented each list five times. This allows to not only replicate traditional serial position effects, but also to see how learning influences positional effects. Three conditions were compared: \textit{serial-constant},\textit{ free-constant}, and \textit{free-varied}. In the \textit{free-varied} condition, list items were randomized between presentations and recalled in any order. In the \textit{free-constant} condition, items instead maintained fixed positions but were recalled with the same free-recall instructions as the \textit{free-varied} condition. In the \textit{serial-constant} condition, item order was fixed and participants were instructed to recall items in the order that they were received.

Participant responses were scored according to their overall accuracy and serial position curve, in addition to analyzing patterns of errors using \textit{conditional response probabilities} (CRP). Serial position curves measure recall accuracy as a function of each item's position in the list. Typically, recall is best for items presented early in the list (the \textit{primacy effect}) and the last several items presented (the \textit{recency effect}; \citeNP{steiner1989immediate}). A hallmark of serial recall is a relatively stronger primacy effect compared to free recall, although the recency effect can be attenuated by introducing a delay between list presentation and subsequent recall.  CRP quantifies the probability of recalling an item as a function of its lag distance from the previously recalled item, providing a measure of contiguity effects. These effects are characterized by a higher likelihood of recalling items close in serial order. Specifically, it shows the probability of recalling item $i + lag$ after recalling item $i$ \cite{kahana2002age}. For example, skipping one item in the list yields a CRP $lag = 1$ for that item, but then recalling the subsequent item correctly would be deemed correct by CRP, whereas returning to the single skipped item would be considered $lag = -1$. 

Stimuli consisted of 19 non-repeating words presented verbally every 1500 ms. Participant responses were scored for overall accuracy, serial position curves, and conditional response probabilities (CRP). Serial position curves capture recall accuracy as a function of list position, typically exhibiting primacy and recency effects \cite{steiner1989immediate}. Serial recall generally produces a stronger primacy effect than free recall, whereas the recency effect diminishes with delayed recall \cite{kahana2002age}.

Results from this study revealed robust primacy and recency effects, with relatively stronger primacy and CRP effects in the \textit{serial-constant} condition. In the two free-recall conditions performance was characterized by a pronounced recency advantage. Across successive presentations of the same list, both primacy and recency became less extreme as overall accuracy increased, gradually flattening the serial position curves while improving recall performance. Contiguity effects were most pronounced from $serialConstant \Rightarrow freeConstant \Rightarrow freeVaried$ where errors became more forward-asymmetric (see Figure \ref{fig:CRP}).

\begin{figure}[htb]
    \centering
    \includegraphics[width=1\linewidth]{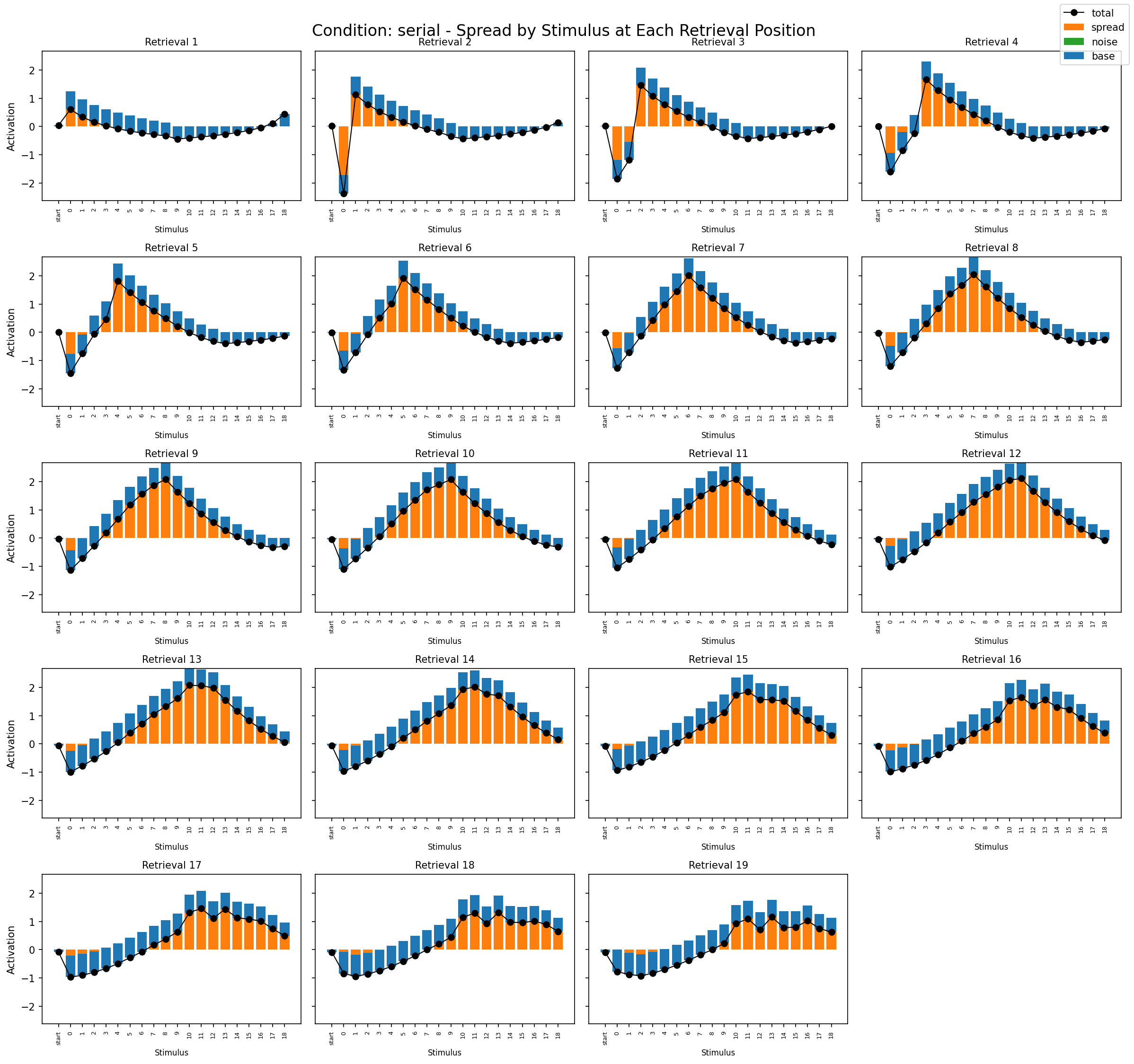}
    \caption{Example of spreading activation during recall phase of serial memory condition. There is an essential tension between the contribution of spreading activation and base-level activation causing run-away activation around the 8th list item where the contextual spread from previous elements overwhelms the inhibition from the recently retrieved item.}
    \label{fig:spreading}
    \end{figure}

\citeA{klein2005comparative} and \citeA{thomson2015account} both interpret these results as supportive of an associative explanation for contextual memory. Primacy effects are driven via rehearsal and clearer context at the start of the list (see Figure \ref{fig:spreading}), with recency effects are driven via the decaying nature of activation dynamics.  Forward-asymmetric contiguity effects - and how they change over time - is driven by associative strengthening and spreading activation via contextual priming. 

\subsection{Impact on Serial Memory Literature}
\citeA{klein2005comparative} provides a relatively novel test case for contemporary theories of list memory, particularly because most existing models have been evaluated on a narrower set of phenomena (e.g., single-presentation studies). More recently, \citeA{adrogue2024multitrial} have further advocated for multi-trial serial memory studies. Generally speaking, most context-based accounts were not implemented with the goal of capturing the joint impact of learning, free- and serial recall together. 

Context-based accounts such as the temporal context model (TCM; also termed the context maintenance and retrieval model, CMR) characterize memory as associations between items and evolving contextual states; recalling an item reinstates its context, biasing retrieval toward items with similar temporal context and jointly accommodating recency and asymmetric contiguity effects \cite{howard1999contextual,polyn2009context}. Converging neural evidence indicates that such temporally-related context representations are reinstated during recall, and that experimental manipulation of temporal context in medial temporal lobe structures modulates contiguity in a manner consistent with these models \cite{manns2011oscillatory,elkalliny2019changing}. Although TCM/CMR has been argued to extend naturally to serial recall, explicit serial-order implementations and quantitative fits have not yet been fully articulated, and current formulations do not directly address how serial position and contiguity functions should evolve over repeated list presentations. Thomson et al.'s \citeyear{thomson2015account} associative-learning framework can be viewed as complementary: by explicitly modeling how associative strengths change with repeated exposure, it instantiates one concrete route by which theories in the TCM/CMR family could, in principle, be extended to accommodate multi-presentation recall data.

The Start–End Model (SEM) explains serial recall by positing implicit \textit{start} and \textit{end} markers, along with positional tokens that encode spatiotemporal location \cite{henson1998short}. One challenge of this theory is the fact that the participant does not necessarily know when the list will end and items outside of working memory would need to have access to the \textit{end} markers. This limits its psychological plausibility. 

SIMPLE offers a unified account of serial and free recall by grounding retrieval in the temporal distinctiveness of items, with optional extensions to other dimensions such as semantic distinctiveness \cite{brown2007temporal}. Empirical applications have primarily emphasized fits to primacy and recency, and, in its standard instantiations, SIMPLE does not appear to reproduce the full pattern of asymmetric contiguity effects or their systematic modulation across repeated list presentations. \citeA{thomson2015account}'s model shares SIMPLE’s commitment to temporal factors, but replaces its more generic distinctiveness construct with a more explicit account of associative strengthening which naturally yields asymmetric contiguity across repetitions \cite{thomson2015account}.

\section{Model Description}
The present model is instantiated using the ACT-Up core architecture supplemented with the contextual and event modules described above. While \citeA{thomson2015account} required declarative FINSTs to achieve their performance, the present model does not use this mechanism as it is psychologically implausibe. Instead the present model uses the \textit{capped-spread} and \textit{actr-decay-activation} functions for associative learning and working memory, respectively. It also uses the \textit{spacing-effect} function for memory retrieval to avoid run-away base-level activation during rehearsal. Most parameters are set to act-r default as appropriate or the same as prior studies (\textit{redacted for peer review}; base-level learning = .5, noise = .25, spreading activation = 4, spacing-delay = 1, and contextual memory decay = .35). 

For a trial, the model assumes a 'start' concept in working memory that serves as the initial context in learning the list (someone analogous to the Start-End Model's more plausible 'start' signal). There is no prior knowledge of the stimuli in the study and no prior experience (i.e., no prior memory activation or associative links between items). Upon perceiving an item the model attempts to rehearse it, although the rapid pace of the experiment means there are few rehearsals. When a chunk is retrieved it is put into working memory, and when that activation has decayed, the item is purged. When the model perceives a new item it is retrieved, rehearsed if the same item is still perceived, and is added to working memory. Associations are strengthened from the preceding item(s) to the newly recalled item. Additionally, associations from more distant items are attenuated by their temporal distance.

After all stimuli are presented the model is prompted to recall an element of the list, driven solely by chunk activation and spreading activation from the items in working memory. Retrievals proceed until the complete list has been recalled or until a recall request fails, at which point the presentation is considered complete. If the model tries to recall the same item again, it tries to retrieve again. The only difference between the free and serial recall tasks is that the model first places the 'start' concept in working memory to prime the first few elements in the list. From that point, preliminary retrievals continue to be sources of activation, creating a robust chain of contextual information to prime subsequent recall.

\subsection{Preliminary Model Results}
\begin{figure}[!htb]
    \centering
    \includegraphics[width=1\linewidth]{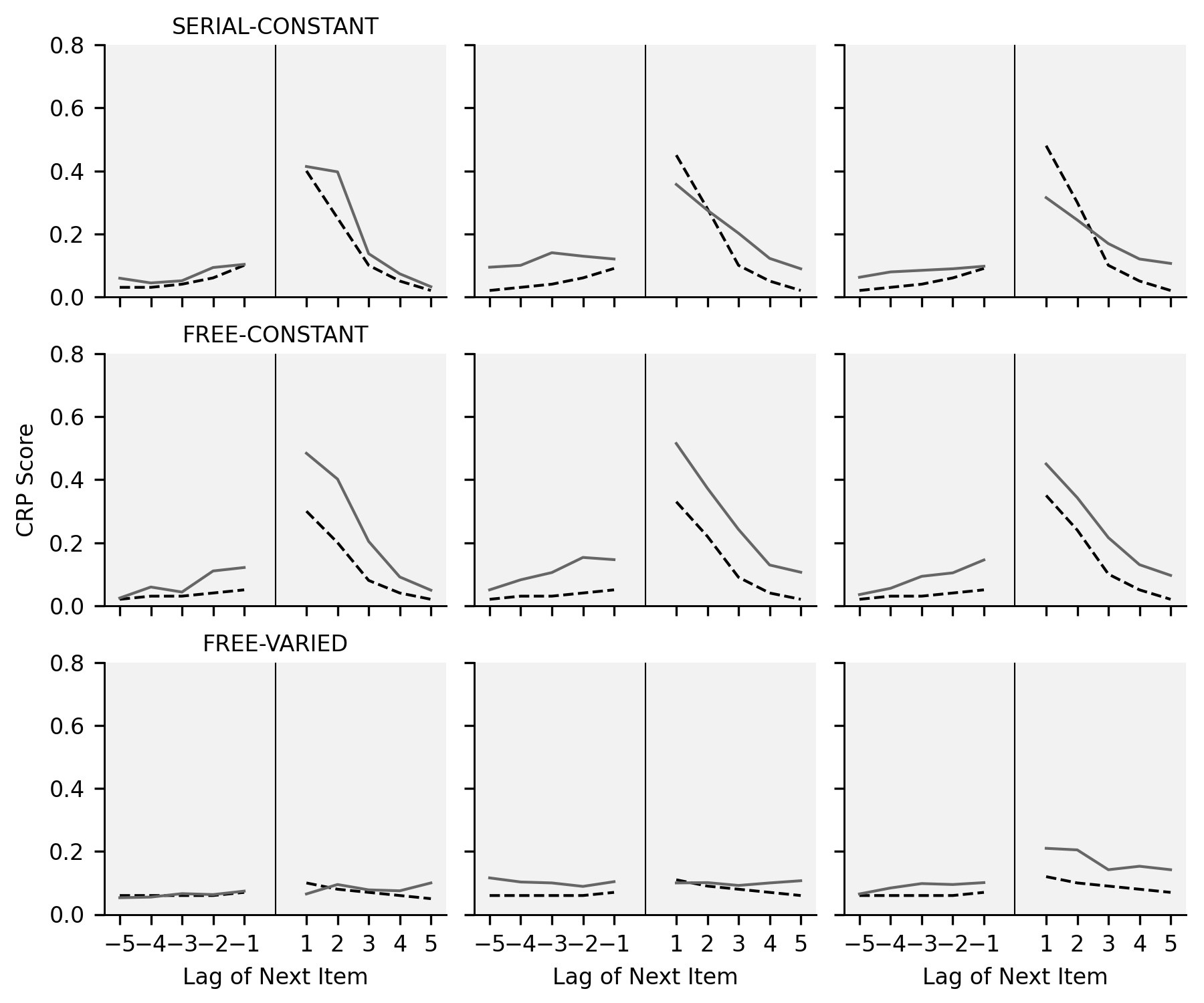}
    \caption{Conditional Response Probabilities across \textit{serial-constant}, \textit{free-constant}, and \textit{free-varied}, showing the probability of recalling item $i + lag$ after $i$. Paneled from left to right are the results for presentations 1, 3, and 5, respectively.}
    \label{fig:CRP}
    \vspace{12pt}
\end{figure}

The model was run for 100 iterations of each trial, and accurately predicts conditional response probabilities (CRPs; $r^2 >.9$) across the serial and free-constant conditions and across repeated presentations (see Figure \ref{fig:CRP}), with improving fits for the free-varied condition over presentations. The spacing-effect addressed prior concerns balancing the contributions of base-level and spreading activation, although it mitigated priming effects due to the tapering effect of reducing the impact of rehearsal during list presentation. The decaying base-level activation creates a recency effect, while primacy is enhanced by associative priming due to the \textit{'start'} concept in working memory serving as context. It is important to note that the 'start' concept is not in the model's memory and only serves as a strategic stand-in for the kinds of priming one would expect from the \textit{Temporal Context} and \textit{Start-End} Models. The asymmetric associative spread clearly explains Conditional Response Probabilities. In the \textit{Free-Varied} condition this asymmetry attenuates over multiple presentations due to randomization of the presented items. 

\subsection{Model Scalability}
The event handler adds negligible overhead. Scheduling an event re-sorts the queue $Q$, which is worst-case $O(Q \log Q)$.
\vspace{-9pt}
\paragraph{Context window.}
Retrievals in ACT-UP already scan all $N$ chunks in memory; the number of active chunks in context $M$ are updated, where $M\ll N$. This adds a negligible factor $O(M)$, which can be viewed as a constant-factor slowdown. 
\vspace{-9pt}
\paragraph{Associative learning.}
Default associative learning operates on the active context elements and their outgoing links. Let $d$ denote the average number of associative links per context source. Updating associations from all $M$ context sources to a single target chunk costs $\Delta T_{\text{assoc}} = O(M d)$ per update, independent of $N$ and negligible for $M, d \ll N$.

\section{Discussion}
This paper serves a triple-purpose: it describes 1) a scalable contextual memory implementation for the ACT-Up cognitive architecture integrating working memory with associative learning without significantly impacting run-time, 2) an event handling module to rapidly prototype psychology experiments with instructions on how to use generative AI to extract study code from academic papers, and 3) a case study on modeling the multiple-presentation serial recall study described in \citeA{klein2005comparative} integrating the aforementioned capabilities with generative AI prompting to rapidly prototype learning mechanisms.  The model achieves comparable performance for \textit{conditional response probabilities} over prior modeling efforts without using explicit non-psychologically plausible strategies such as only rehearsing the first two list items and using declarative FINSTs to avoid pathological retrieval-loops \cite{thomson2015account}. 

Beyond providing a practical toolkit for rapidly prototyping cognitive mechanisms, the provided contextual memory framework provides a viable bridge between extant theories of serial memory (TCM/CMR, SIMPLE, and SEM) for explaining human behavior in multi-presentation serial memory paradigms. At the same time, challenges to the integration of multiple mechanisms together indicate that further research is required to better integrate base-level and associative activation into a stable and cohesive activation function. This toolkit provides the means to rapidly prototype candidate mechanisms, with a 100 repetition model run taking approximately 30 seconds, of which logging activations is a substantial portion of that overhead. 

Further efforts will include developing automated tools to inspect activations, develop and integrate additional mechanisms (e.g., structure mapping \cite{thomson2025cognitive} and similarity learning), and validate each mechanism against human data. We additionally plan to investigate the degree to which the next-generation of generative AI can use the ACT-Up architecture, as a single Lisp file, to develop basic models in addition to extracting the required events to model a given task directly from a research paper.

\section{Acknowledgments}
This research was funded, in part, by the Advanced Research Projects Agency for Health (ARPA-H) from work originally funded by the Office of the Secretary of Defense / Assistant Secretary of Defense for Research and Engineering and the Office of Naval Research. The views and conclusions contained in this document are those of the authors and should not be interpreted as representing the official policies, either expressed or implied, of the U.S. Government, the Office of Naval Research, or the U.S. Navy.

\bibliographystyle{apacite}

\setlength{\bibleftmargin}{.125in}
\setlength{\bibindent}{-\bibleftmargin}

\bibliography{ICCM_Template}

@article{manns2011oscillatory,
  author  = {J. R. Manns and M. W. Howard and H. Eichenbaum},
  title   = {Oscillatory patterns in the temporal lobe reveal context reinstatement during memory retrieval},
  journal = {Proceedings of the National Academy of Sciences},
  year    = {2011},
  volume  = {108},
  number  = {31},
  pages   = {12893--12897},
  doi     = {10.1073/pnas.1015174108}
}

@article{elkalliny2019changing,
  title={Changing temporal context in human temporal lobe promotes memory of distinct episodes},
  author={El-Kalliny, Mostafa M and Wittig Jr, John H and Sheehan, Timothy C and Sreekumar, Vishnu and Inati, Sara K and Zaghloul, Kareem A},
  journal={Nature communications},
  volume={10},
  number={1},
  pages={203},
  year={2019},
  publisher={Nature Publishing Group UK London}
}

@article{adrogue2024multitrial,
  title={Multitrial free recall for evaluating memory.},
  author={Adrogue, RT and Herz, N and Halpern, DJ and Tracy, J and Kahana, MJ},
  journal={Neuropsychology},
  volume={38},
  number={1},
  pages={58},
  year={2024},
  publisher={American Psychological Association}
}

@article{henson1998short,
  title={Short-term memory for serial order: The start-end model},
  author={Henson, Richard NA},
  journal={Cognitive psychology},
  volume={36},
  number={2},
  pages={73--137},
  year={1998},
  publisher={Elsevier}
}

@article{peterson1959short,
  title={Short-term retention of individual verbal items.},
  author={Peterson, Lloyd and Peterson, Margaret Jean},
  journal={Journal of experimental psychology},
  volume={58},
  number={3},
  pages={193},
  year={1959},
  publisher={American Psychological Association}
}

@misc{lupyan2026unreasonableeffectivenesspatternmatching,
      title={The unreasonable effectiveness of pattern matching}, 
      author={Gary Lupyan and Blaise Agüera Arcas},
      year={2026},
      eprint={2601.11432},
      archivePrefix={arXiv},
      primaryClass={cs.CL},
      url={https://arxiv.org/abs/2601.11432}, 
}

@article{anderson1991reflections,
  title={Reflections of the environment in memory},
  author={Anderson, John R and Schooler, Lael J},
  journal={Psychological science},
  volume={2},
  number={6},
  pages={396--408},
  year={1991},
  publisher={SAGE Publications Sage CA: Los Angeles, CA}
}

@article{pavlik2005practice,
  title={Practice and forgetting effects on vocabulary memory: An activation-based model of the spacing effect},
  author={Pavlik, Philip I and Anderson, John R},
  journal={Cognitive science},
  volume={29},
  number={4},
  pages={559--586},
  year={2005},
  publisher={Wiley Online Library}
}

@article{thomson2025cognitive,
  title={Cognitive models of influence dynamics in a conformity simulation},
  author={Thomson, Robert and Lebiere, Christian},
  journal={Computational and Mathematical Organization Theory},
  volume={31},
  number={4},
  pages={323--343},
  year={2025},
  publisher={Springer}
}

@inproceedings{reitter2010accountable,
  title={Accountable modeling in ACT-UP, a scalable, rapid-prototyping ACT-R implementation},
  author={Reitter, David and Lebiere, Christian},
  booktitle={Proceedings of the 10th international conference on cognitive modeling (iccm)},
  pages={199--204},
  year={2010},
  organization={Citeseer}
}

@article{thomson2017account,
  title={An account of interference in associative memory: Learning the fan effect},
  author={Thomson, Robert and Harrison, Anthony M and Trafton, J Gregory and Hiatt, Laura M},
  journal={Topics in Cognitive Science},
  volume={9},
  number={1},
  pages={69--82},
  year={2017},
  publisher={Wiley Online Library}
}

@inproceedings{thomson2015account,
  title={An Account of Associative Learning in Memory Recall.},
  author={Thomson, Robert and Pyke, Aryn and Hiatt, Laura M and Trafton, J Greg},
  booktitle={Proceedings of the Annual Meeting of the Cognitive Science Society},
  volume={37},
  pages={2386--2391},
  year={2015}
}

@inproceedings{thomson2014extending,
  title={Extending the influence of contextual information in ACT-R using buffer decay},
  author={Thomson, Robert and Bennati, Stefano and Lebiere, Christian},
  booktitle={Proceedings of the annual meeting of the cognitive science society},
  volume={36},
  number={36},
  year={2014}
}

@inproceedings{thomson2013constraining,
  title={Constraining Bayesian inference with cognitive architectures: An updated associative learning mechanism in ACT-R},
  author={Thomson, Robert and Lebiere, Christian},
  booktitle={Proceedings of the annual meeting of the cognitive science society},
  volume={35},
  number={35},
  year={2013}
}

@article{stocco2024integrated,
  title={An integrated computational framework for the neurobiology of memory based on the ACT-R declarative memory system},
  author={Stocco, Andrea and Rice, Patrick and Thomson, Robert and Smith, Briana and Morrison, Don and Lebiere, Christian},
  journal={Computational Brain \& Behavior},
  volume={7},
  number={1},
  pages={129--149},
  year={2024},
  publisher={Springer}
}

@inproceedings{lebiere2015functional,
  title={Functional cognitive models of malware identification},
  author={Lebiere, Christian and Bennati, Stefano and Thomson, Robert and Shakarian, Paulo and Nunes, Eric},
  booktitle={13th International Conference on Cognitive Modeling, ICCM 2015},
  pages={90--95},
  year={2015},
  organization={University of Groningen}
}

@article{klein2005comparative,
  title={A comparative analysis of serial and free recall},
  author={Klein, Krystal A and Addis, Kelly M and Kahana, Michael J},
  journal={Memory \& Cognition},
  volume={33},
  number={5},
  pages={833--839},
  year={2005},
  publisher={Springer}
}

@techreport{anderson2020associative,
    author = {Anderson, John R},
    title = {Revisiting Associative Learning},
    institution = {ACT-R Workshop Proceedings},
    address = {Pittsburgh, PA},
    year = {2020}
}

@inproceedings{yang2020pyactup,
    author = {Yang, Cher and Morrison, Don and Stocco, Andrea and Orr, Mark and Lebiere, Christian},
    title = {An Expanded Set of Declarative Memory Functionalities in PyACTUp, a Python Implementation of ACT-UP’s Accountable Modeling},
    booktitle = {Proceedings of the MathPsych/ICCM Annual Conference},
    year = {2020}
}

@inproceedings{kelly2015holographic,
  title={Holographic declarative memory and the fan effect: A test case for a new memory module for ACT-R},
  author={Kelly, Matthew A and Kwok, Kam and West, Robert L},
  booktitle={Proceedings of the 13th International Conference on Cognitive Modeling. Groningen, the Netherlands: University of Groningen},
  pages={148--153},
  year={2015}
}

@article{kahana2002age,
  title={Age dissociates recency and lag recency effects in free recall.},
  author={Kahana, Michael J and Howard, Marc W and Zaromb, Franklin and Wingfield, Arthur},
  journal={Journal of Experimental Psychology: Learning, Memory, and Cognition},
  volume={28},
  number={3},
  pages={530},
  year={2002},
  publisher={American Psychological Association}
}

@article{steiner1989immediate,
  title={Immediate and delayed primacy and recency effects in performance evaluation.},
  author={Steiner, Dirk D and Rain, Jeffrey S},
  journal={Journal of Applied Psychology},
  volume={74},
  number={1},
  pages={136},
  year={1989},
  publisher={American Psychological Association}
}

@article{howard1999contextual,
  title={Contextual variability and serial position effects in free recall.},
  author={Howard, Marc W and Kahana, Michael J},
  journal={Journal of Experimental Psychology: Learning, Memory, and Cognition},
  volume={25},
  number={4},
  pages={923},
  year={1999},
  publisher={American Psychological Association}
}

@article{brown2007temporal,
  title={A temporal ratio model of memory.},
  author={Brown, Gordon DA and Neath, Ian and Chater, Nick},
  journal={Psychological review},
  volume={114},
  number={3},
  pages={539},
  year={2007},
  publisher={American Psychological Association}
}

@article{polyn2009context,
  title={A context maintenance and retrieval model of organizational processes in free recall.},
  author={Polyn, Sean M and Norman, Kenneth A and Kahana, Michael J},
  journal={Psychological review},
  volume={116},
  number={1},
  pages={129},
  year={2009},
  publisher={American Psychological Association}
}

\end{document}